\documentclass[12pt,english,aps,superscriptaddress,preprintnumbers]{revtex4-1}
\usepackage[T1]{fontenc}
\usepackage[latin9]{inputenc}
\usepackage{float}
\usepackage{mathrsfs}
\usepackage{amsmath}
\usepackage{amssymb}
\usepackage{graphicx}
\usepackage{esint}

\makeatletter

\usepackage{mathrsfs}

\@ifundefined{textcolor}{}{%
 \definecolor{BLACK}{gray}{0}
 \definecolor{WHITE}{gray}{1}
 \definecolor{RED}{rgb}{1,0,0}
 \definecolor{GREEN}{rgb}{0,1,0}
 \definecolor{BLUE}{rgb}{0,0,1}
 \definecolor{CYAN}{cmyk}{1,0,0,0}
 \definecolor{MAGENTA}{cmyk}{0,1,0,0}
 \definecolor{YELLOW}{cmyk}{0,0,1,0}
}

\usepackage{babel}

\makeatother

\usepackage{babel}
\begin{document}
\title{Thermodynamics of the system of massive Dirac fermions in a uniform
magnetic field}
\author{Ren-Hong Fang}
\affiliation{Key Laboratory of Particle Physics and Particle Irradiation (MOE),
Institute of Frontier and Interdisciplinary Science, Shandong University,
Qingdao, Shandong 266237, China}
\author{Ren-Da Dong }
\affiliation{Institute of Particle Physics and Key Laboratory of Quark and Lepton
Physics (MOS), Central China Normal University, Wuhan 430079, China}
\author{De-Fu Hou }
\email{Corresponding authors. houdf@mail.ccnu.edu.cn; sunbd@sdu.edu.cn}

\affiliation{Institute of Particle Physics and Key Laboratory of Quark and Lepton
Physics (MOS), Central China Normal University, Wuhan 430079, China}
\author{Bao-Dong Sun}
\email{Corresponding authors. houdf@mail.ccnu.edu.cn; sunbd@sdu.edu.cn}

\affiliation{Key Laboratory of Particle Physics and Particle Irradiation (MOE),
Institute of Frontier and Interdisciplinary Science, Shandong University,
Qingdao, Shandong 266237, China}
\begin{abstract}

We construct the grand partition function of the system of massive
Dirac fermions in a uniform magnetic field from Landau levels, through
which all thermodynamic quantities can be obtained. Making use of
the Abel-Plana formula, these thermodynamic quantities can be expanded
as power series with respect to the dimensionless variable $b=2eB/T^{2}$.
The zero-field magnetic susceptibility is expanded at zero mass, and
the leading order term is logarithmic. We also calculate scalar, vector
current, axial vector current and energy-momentum tensor of the system
through ensemble average approach. Mass correction to chiral separation
effect is discussed. For massless chiral fermions, our results recover
the chiral magnetic effect for right- and left-handed fermions, as
well as chiral separation effect.
\end{abstract}
\maketitle

\section{Introduction}

Since the invention of the Dirac equation \citep{Dirac:1928hu} and
the discovery of the positron \citep{Anderson:1933mb}, the properties
of the fermion system under an electromagnetic field have been studied
for years. Under the condition of strong electric field at order $1$
$\mathrm{MeV}^{2}$, fermion pairs can emerge from the vacuum, which
is called Schwinger pair production and changes the structure of Maxwell
equation in vacuum \citep{Heisenberg:1935qt,Weisskopf:1936hya}. For
the vacuum quantum field theory, the proper time approach was developed
for fermion systems interacting with an external electromagnetic field
at zero temperature \citep{Schwinger:1951,Salam:1974xe,Blau:1988iz}.
For the finite temperature fermion systems, the imaginary-time approach
is extensively used to study the thermodynamics of the systems \citep{Dolan:1973qd,Shuryak:1980tp,Kapusta:2006}.
The effective action for the system of massive fermions in the presence
of a static background magnetic field was obtained by the computational
techniques of the imaginary time formalism and the proper time method
\citep{Cangemi:1996tp}. Wigner function is a useful tool to study
the properties of the thermodynamics and hydromechanics of the fermion
system not only for the background of a pure magnetic field \citep{Hebenstreit:2010vz,Sheng:2017lfu,Sheng:2018jwf}
but also for a general electromagnetic field \citep{Vasak:1987um,Gao:2012ix,Chen:2012ca,Hidaka:2016yjf,Gao:2018wmr,Yang:2020mtz,Guo:2020zpa},
which automatically combines the effects of both the finite temperature
and the electromagnetic field in a covariant and gauge-invariant way. 
The strong magnetic field produced by the rapidly rotating compact stars 
may have great impact on the state of them \citep{Itokazu:2018lij,Reisenegger:2013faa,Islam:2018lmy}. 
Based on the theoretical approaches, the influence of the magnetic
field on the phenomenology in high energy heavy ion collisions, such
as chiral magnetic effect \citep{Kharzeev:2007jp,Fukushima:2008xe,Gao:2012ix,Feng:2018tpb,Shi:2018izg,STAR:2019bjg,Liang:2020sgr,Gao:2020vbh,Liu:2020ymh},
chiral separation effect \citep{Son:2004tq,Metlitski:2005pr,Lin:2018aon},
magnetic catalysis \citep{Bali:2013esa,Mao:2015kva,Mao:2018wqo,Ballon-Bayona:2020xtf},
ect., is being widely researched nowadays. In recent years, the method
of lattice calculation is used to study the magnetic properties of
QCD matter \citep{Bali:2020bcn,Ding:2020pao,Buividovich:2021fsa}.

In this article we study the thermodynamics of the system of massive
Dirac fermions in a uniform magnetic field. Dirac fermion means that
the wavefunction of the fermion satisfies the Dirac equation. Firstly
we construct the grand partition function of the system from the Landau
levels for a single massive fermion in a uniform magnetic field. This
is different from the imaginary time formalism and the proper time
method used in \citep{Cangemi:1996tp}, where the authors calculated
the effective Lagrangian of the fermion system and the vacuum term
was naturally included. In this article, we have ignored the vacuum
term in the grand partition function, which is equivalent to the normal
ordering description of ensemble average approach \citep{Vasak:1987um,Dong:2020zci}
and different from the un-normal ordering description in recent articles
\citep{Sheng:2017lfu,Gao:2019zhk,Yang:2020mtz}. From the grand partition
function, we can obtain all thermodynamic quantities, which can be
expanded as power series of the dimensionless variable $b=2eB/T^{2}$
making use of the Abel-Plana formula. It is worth pointing out that
all thermodynamic quantities are analytic at $b=0$. However, for
the massless fermion case, the expansions of some thermodynamic quantities
contain a singular term $\ln b^{2}$ as discussed in a recent article
by some of the authors \citep{Fang:2020efk}. To avoid possible confusion,
we point out that, in this article the singular log term does not
mean the infinity of the physical quantity itself, but means the infinity
of the derivative of the physical quantity with respect to the magnetic
field at $B=0$. We expand the zero-field magnetic susceptibility
with respect to the dimensionless mass parameter $c=m/T$, which is
consistent with the result by the proper time method \citep{Bali:2020bcn}.
In the cases of high and low temperature, the results for magnetic
susceptibility are qualitatively consistent with lattice calculation.
The scalar, vector current, axial vector current and energy-momentum
tensor of the system are calculated through ensemble average approach.
For the axial vector current, we expand the coefficient of chiral
separation effect at $c=0$ to arbitrary orders, and our results recover
the chiral magnetic effect for right- and left-handed fermions in
massless limit.

The rest of this article is organized as follows. In Sec. \ref{sec:Landau-levels},
we list the Landau levels for a single massive Dirac fermion in a
uniform magnetic field. In Sec. \ref{sec:partition function}, all
intensive quantities are expressed by the grand partition function.
In Sec. \ref{sec:Expansion}, making use of Abel-Plana formula, we
expand all intensive quantities as power series of the dimensionless
variable $b=2eB/T^{2}$. In Sec. \ref{sec:Zero-field}, the zero-field
magnetic susceptibility is expanded with respect to the dimensionless
mass parameter $c=m/T$. In Sec. \ref{sec:trace}, we calculate the
scalar, and energy-momentum tensor of the system through ensemble
average approach. In Sec. \ref{sec:CSE}, mass correction to chiral
separation effect is discussed. This article is summarized in Sec.
\ref{sec:Summary}.

Throughout this article we adopt natural units where $\hbar=c=k_{B}=1$.
Metric tensor is $g^{\mu\nu}=\mathrm{diag}\,(+1,-1,-1,-1)$. Heaviside-Lorentz
convention is used for electromagnetism.

\section{Landau levels for Dirac equation in a uniform magnetic field}

\label{sec:Landau-levels}In this article, we use chiral representation
for Dirac gamma matrixes, where the explicit forms of $\gamma^{\mu}$
and $\gamma^{5}$ are
\begin{equation}
\gamma^{\mu}=\left(\begin{array}{cc}
0 & \sigma^{\mu}\\
\bar{\sigma}^{\mu} & 0
\end{array}\right),\ \ \ \ \gamma^{5}=i\gamma^{0}\gamma^{1}\gamma^{2}\gamma^{3}=\left(\begin{array}{cc}
-1 & 0\\
0 & 1
\end{array}\right)\label{eq:71a}
\end{equation}
where $\sigma^{\mu}=(1,\boldsymbol{\sigma})$, $\bar{\sigma}^{\mu}=(1,-\boldsymbol{\sigma})$,
and $\boldsymbol{\sigma}=(\sigma^{1},\sigma^{2},\sigma^{3})$ are
the three Pauli matrixes. Dirac equation for a massive fermion in
a uniform magnetic field $\boldsymbol{B}=B\boldsymbol{e}_{z}$ is
\begin{equation}
i\frac{\partial\psi}{\partial t}=(i\boldsymbol{\alpha}\cdot\boldsymbol{D}+\gamma^{0}m)\psi,\label{eq:71b-1}
\end{equation}
where $\boldsymbol{\alpha}=\gamma^{0}\boldsymbol{\gamma}$, $\boldsymbol{D}=(-\partial_{x},-\partial_{y}+ieBx,-\partial_{z})$,
and $e$, $m$ are the electric charge and mass of the fermion. We
chose $A^{\mu}=(0,0,Bx,0)$ for the gauge potential. For the sake
of simplicity in following calculations, we will set $eB>0$ throughout
this article. The range $eB>0$ can be analytically continued to the
range $eB<0$ for all results in this article.

The solutions of the energy eigenvalue equation $(i\boldsymbol{\alpha}\cdot\boldsymbol{D}+\gamma^{0}m)\psi=E\psi$
give a series of Landau levels and eigenfunctions as follows,
\begin{eqnarray}
n=0,\ \  & E=\lambda E_{0}(k_{z}),\ \  & \psi_{0\lambda}(k_{y},k_{z};\boldsymbol{x})=c_{0\lambda}\left(\begin{array}{c}
\varphi_{0}(\xi)\\
0\\
F_{0\lambda}\varphi_{0}(\xi)\\
0
\end{array}\right)\frac{1}{L}e^{i(yk_{y}+zk_{z})},\label{eq:j1}\\
n>0,\ \  & E=\lambda E_{n}(k_{z}),\ \  & \psi_{n\lambda s}(k_{y},k_{z};\boldsymbol{x})=c_{n\lambda s}\left(\begin{array}{c}
\left(\begin{array}{c}
\varphi_{n}(\xi)\\
iK_{ns}\varphi_{n-1}(\xi)
\end{array}\right)\\
F_{n\lambda s}\left(\begin{array}{c}
\varphi_{n}(\xi)\\
iK_{ns}\varphi_{n-1}(\xi)
\end{array}\right)
\end{array}\right)\frac{1}{L}e^{i(yk_{y}+zk_{z})},\label{eq:519a}
\end{eqnarray}
where $\lambda=\pm1$, $s=\pm1$, $\xi=\sqrt{eB}x-k_{y}/\sqrt{eB}$,
$E_{n}(k_{z})=\sqrt{m^{2}+k_{z}^{2}+2neB}$, $\varphi_{n}(\xi)$ is
the $n$-th harmonic oscillator function along $x$-axis satisfying
$\int_{-\infty}^{\infty}d\xi\varphi_{n}(\xi)\varphi_{m}(\xi)=\sqrt{eB}\delta_{nm}$,
and the coefficients $K_{ns}(k_{z})$, $F_{0\lambda}(k_{z})$, $F_{n\lambda s}(k_{z})$,
$c_{0\lambda}$, $c_{n\lambda s}$ are defined as
\begin{equation}
K_{ns}(k_{z})=\frac{k_{z}+s\sqrt{k_{z}^{2}+2neB}}{\sqrt{2neB}},\label{eq:j2}
\end{equation}
\begin{equation}
F_{0\lambda}(k_{z})=\frac{\lambda E_{0}+k_{z}}{m},\ \ \ F_{n\lambda s}(k_{z})=\frac{\lambda E_{n}-s\sqrt{k_{z}^{2}+2neB}}{m},\label{eq:j3}
\end{equation}
\begin{equation}
|c_{0\lambda}|^{2}=\frac{1}{1+F_{0\lambda}^{2}},\ \ |c_{n\lambda s}|^{2}=\frac{1}{(1+F_{n\lambda s}^{2})(1+K_{ns}^{2})}.\label{eq:71e}
\end{equation}
The eigenfunctions are set up in a cube with side length $L$, and
its plane wave factor $e^{i(yk_{y}+zk_{z})}$ satisfies periodic boundary
condition, i.e. $k_{y}=2\pi n_{y}/L$, $k_{z}=2\pi n_{z}/L$, $(n_{y},n_{z}\in\mathbb{Z})$.
The center of the harmonic oscillator function $\varphi_{n}(\xi)$
is located at $x_{0}=k_{y}/(eB)$. The condition $0<x_{0}<L$ leads
to $0<n_{y}<eBL^{2}/(2\pi)$. Since the Landau levels do not depend
on $k_{y}$, the degeneracy of each Landau level is $eBL^{2}/(2\pi)$.
The calculation details of Landau levels can be found in \citep{Dariescu:2015kva,Sheng:2017lfu,Sheng:2019ujr}.

\section{Grand partition function and thermodynamic quantities}

\label{sec:partition function}Now we consider a system of massive
Dirac fermions in a uniform magnetic field $\boldsymbol{B}=B\boldsymbol{e}_{z}$,
where the interaction among fermions is ignored. This system is in
equilibrium with a reservoir, which keeps constant temperature $T$
and constant chemical potential $\mu$. We can construct the grand
partition function $\ln\Xi$ of this system from Landau levels in
Sec. \ref{sec:Landau-levels} as follows,
\begin{eqnarray}
\ln\Xi & = & \sum_{k_{y},k_{z}}\bigg[\ln\left(1+e^{\beta\mu-\beta\sqrt{m^{2}+k_{z}^{2}}}\right)+\ln\left(1+e^{-\beta\mu-\beta\sqrt{m^{2}+k_{z}^{2}}}\right)\bigg]\nonumber \\
 &  & +2\sum_{n=1}^{\infty}\sum_{k_{y},k_{z}}\bigg[\ln\left(1+e^{\beta\mu-\beta\sqrt{2neB+m^{2}+k_{z}^{2}}}\right)+\ln\left(1+e^{-\beta\mu-\beta\sqrt{2neB+m^{2}+k_{z}^{2}}}\right)\bigg],\label{eq:5.1d}
\end{eqnarray}
where $\beta=1/T$. We have ignored the vacuum term in $\ln\Xi$,which is equivalent to the normal ordering description of ensemble
average approach \citep{Vasak:1987um,Dong:2020zci}.  Here we are only interested in the medium induced correction, and neglect the vacuum contribution to the grand partition function.  Actually  the vacuum term would give an  temperature-independent  ultraviolet divergence  which can be regularized by conventional re-normalization  method  at zero temperature.  The summations
over $k_{y}$ and $k_{z}$ in Eq. (\ref{eq:5.1d}) can be replaced
by the degeneracy factor $eBL^{2}/(2\pi)$ and the integral $(L/2\pi)\int dk_{z}$
respectively.

The thermodynamic quantities of the system, such as particle number
$N=Vn$, energy $U=V\varepsilon$, pressure $p$, entropy $S=Vs$
and magnetization intensity $M$, can be derived from $\ln\Xi$ in
the following,
\begin{equation}
N=\frac{\partial}{\partial a}\ln\Xi,\label{eq:512e}
\end{equation}
\begin{equation}
U=-\frac{\partial}{\partial\beta}\ln\Xi,\label{eq:512f}
\end{equation}
\begin{equation}
p=\frac{1}{\beta}\frac{\partial}{\partial V}\ln\Xi,\label{eq:512g}
\end{equation}
\begin{equation}
S=\ln\Xi+\beta U-aN,\label{eq:512h}
\end{equation}
\begin{equation}
M=\frac{1}{\beta}\frac{\partial}{\partial B}\bigg(\frac{\ln\Xi}{V}\bigg).\label{eq:512h-1}
\end{equation}
where $V=L^{3}$ is the volume of the system. After introducing an
intensive quantity $g(a,b,c)\equiv(\beta^{3}\ln\Xi)/L^{3}$ and three
dimensionless variables, $a=\beta\mu$, $b=2eB\beta^{2}$, $c=\beta m$,
the grand partition function in Eq. (\ref{eq:5.1d}) becomes
\begin{eqnarray}
g(a,b,c) & = & \frac{b}{4\pi^{2}}\int_{0}^{\infty}ds\bigg[\ln(1+e^{a-\sqrt{s^{2}+c^{2}}})+\ln(1+e^{-a-\sqrt{s^{2}+c^{2}}})\bigg]\nonumber \\
 &  & +\frac{b}{2\pi^{2}}\int_{0}^{\infty}ds\sum_{n=1}^{\infty}\bigg[\ln(1+e^{a-\sqrt{nb+s^{2}+c^{2}}})+\ln(1+e^{-a-\sqrt{nb+s^{2}+c^{2}}})\bigg].\label{eq:519b}
\end{eqnarray}
Now the extensive quantities in Eqs. (\ref{eq:512e}-\ref{eq:512h-1})
becomes
\begin{equation}
n=\frac{1}{\beta^{3}}\frac{\partial}{\partial a}g(a,b,c),\label{eq:512i}
\end{equation}
\begin{equation}
\varepsilon=\frac{1}{\beta^{4}}\bigg(3-2b\frac{\partial}{\partial b}-c\frac{\partial}{\partial c}\bigg)g(a,b,c),\label{eq:512j}
\end{equation}
\begin{equation}
p=\frac{1}{\beta^{4}}g(a,b,c),\label{eq:512k}
\end{equation}
\begin{equation}
s=\frac{1}{\beta^{3}}\bigg(4-a\frac{\partial}{\partial a}-2b\frac{\partial}{\partial b}-c\frac{\partial}{\partial c}\bigg)g(a,b,c),\label{eq:512l}
\end{equation}
\begin{equation}
M=\frac{2e}{\beta^{2}}\frac{\partial}{\partial b}g(a,b,c),\label{eq:512m}
\end{equation}
\begin{equation}
\chi=4e^{2}\frac{\partial^{2}}{\partial b^{2}}g(a,b,c),\label{eq:512n}
\end{equation}
\begin{eqnarray}
c_{T} & = & \frac{1}{\beta^{3}}\bigg(12-3a\frac{\partial}{\partial a}-10b\frac{\partial}{\partial b}-6c\frac{\partial}{\partial c}+4b^{2}\frac{\partial^{2}}{\partial b^{2}}+c^{2}\frac{\partial^{2}}{\partial c^{2}}\nonumber \\
 &  & +2ab\frac{\partial^{2}}{\partial a\partial b}+ac\frac{\partial^{2}}{\partial a\partial c}+4bc\frac{\partial^{2}}{\partial b\partial c}\bigg)g(a,b,c),\label{eq:5.1i}
\end{eqnarray}
where $n$, $\varepsilon$, $p$, $s$, $M$, $\chi\equiv\partial M/\partial B$,
$c_{T}\equiv\partial\varepsilon/\partial T$, are particle number
density, energy density, pressure, entropy density, magnetization
intensity, magnetic susceptibility and heat capacity, respectively.

\section{Expansions of intensive quantities at $B=0$}

\label{sec:Expansion} In order to study the analytic behaviors of
all thermodynamic quantities at zero magnetic field, in this section
all thermodynamic quantities will be expanded at $B=0$, i.e. at $b=0$
with $b=2eB\beta^{2}$.

After defining an auxiliary function $f(a,x)$
\begin{equation}
f(a,x)=\ln(1+e^{a-x})+\ln(1+e^{-a-x}),\label{eq:a2}
\end{equation}
$g(a,b,c)$ in Eq. (\ref{eq:519b}) becomes
\begin{equation}
g(a,b,c)=\frac{b}{2\pi^{2}}\int_{0}^{\infty}ds\left[\frac{1}{2}f(a,\sqrt{s^{2}+c^{2}})+\sum_{n=1}^{\infty}f(a,\sqrt{nb+s^{2}+c^{2}})\right].\label{eq:5.1j-1}
\end{equation}
Making use of following Abel-Plana formula \citep{Ni:2003,Butzer:2011}
\begin{equation}
\frac{1}{2}\mathcal{F}(0)+\sum_{n=1}^{\infty}\mathcal{F}(n)=\int_{0}^{\infty}dt\mathcal{F}(t)+i\int_{0}^{\infty}dt\frac{\mathcal{F}(it)-\mathcal{F}(-it)}{e^{2\pi t}-1},\label{eq:5.1k-1}
\end{equation}
the summation over Landau levels in Eq. (\ref{eq:5.1j-1}) can be
transformed into integrations, i.e.
\begin{eqnarray}
g(a,b,c) & = & \frac{1}{2\pi^{2}}\int_{0}^{\infty}ds\int_{0}^{\infty}dtf(a,\sqrt{t+s^{2}+c^{2}})\nonumber \\
 &  & +\frac{b}{2\pi^{2}}\times i\int_{0}^{\infty}ds\int_{0}^{\infty}dt\frac{f(a,\sqrt{itb+s^{2}+c^{2}})-f(a,\sqrt{-itb+s^{2}+c^{2}})}{e^{2\pi t}-1},\label{eq:5.1l-1}
\end{eqnarray}
In Appendix \ref{sec:Expansion-app}, we have expanded $g(a,b,c)$
at $b=0$ as follows,
\begin{equation}
g(a,b,c)=\frac{1}{2\pi^{2}}\int_{0}^{\infty}ds\int_{0}^{\infty}dtf(a,\sqrt{t+s^{2}+c^{2}})-\frac{1}{\pi^{2}}\sum_{n=0}^{\infty}\frac{(4n+1)!!}{(4n+4)!!}\mathscr{B}_{2n+2}C_{2n+1}(a,c)b^{2n+2},\label{eq:519k}
\end{equation}
where $\mathscr{B}_{n}$ are Bernoulli numbers, and $C_{2n+1}(a,c)$
$(n\geqslant0)$ is
\begin{equation}
C_{2n+1}(a,c)=-\frac{1}{(4n+1)!}\int_{0}^{\infty}dy\ln y\frac{d^{4n+2}}{dy^{4n+2}}f(a,\sqrt{y^{2}+c^{2}}).\label{eq:5.1n}
\end{equation}
We can see that $g(a,b,c)$ in Eq. (\ref{eq:519k}) is a power series
of $b^{2}$.

Making use of the power series expansion of $g(a,b,c)$, the intensive
quantities in Eqs. (\ref{eq:512i})-(\ref{eq:5.1i}) become
\begin{eqnarray}
n\beta^{3} & = & \frac{1}{2\pi^{2}}\int_{0}^{\infty}ds\int_{0}^{\infty}dt\frac{\partial}{\partial a}f(a,\sqrt{t+s^{2}+c^{2}})\nonumber \\
 &  & -\frac{1}{\pi^{2}}\sum_{n=0}^{\infty}\frac{(4n+1)!!}{(4n+4)!!}\mathscr{B}_{2n+2}\frac{\partial}{\partial a}C_{2n+1}(a,c)b^{2n+2},\label{eq:519c}
\end{eqnarray}

\begin{eqnarray}
\varepsilon\beta^{4} & = & \frac{1}{2\pi^{2}}\int_{0}^{\infty}ds\int_{0}^{\infty}dt\left(3-c\frac{\partial}{\partial c}\right)f(a,\sqrt{t+s^{2}+c^{2}})\nonumber \\
 &  & +\frac{1}{\pi^{2}}\sum_{n=0}^{\infty}\frac{(4n+1)!!}{(4n+4)!!}\mathscr{B}_{2n+2}\left(4n+1+c\frac{\partial}{\partial c}\right)C_{2n+1}(a,c)b^{2n+2},\label{eq:519d}
\end{eqnarray}
\begin{eqnarray}
p\beta^{4} & = & \frac{1}{2\pi^{2}}\int_{0}^{\infty}ds\int_{0}^{\infty}dtf(a,\sqrt{t+s^{2}+c^{2}})\nonumber \\
 &  & -\frac{1}{\pi^{2}}\sum_{n=0}^{\infty}\frac{(4n+1)!!}{(4n+4)!!}\mathscr{B}_{2n+2}C_{2n+1}(a,c)b^{2n+2},\label{eq:519l}
\end{eqnarray}
\begin{eqnarray}
s\beta^{3} & = & \frac{1}{2\pi^{2}}\int_{0}^{\infty}ds\int_{0}^{\infty}dt\left(4-a\frac{\partial}{\partial a}-c\frac{\partial}{\partial c}\right)f(a,\sqrt{t+s^{2}+c^{2}})\nonumber \\
 &  & +\frac{1}{\pi^{2}}\sum_{n=0}^{\infty}\frac{(4n+1)!!}{(4n+4)!!}\mathscr{B}_{2n+2}\left(4n+a\frac{\partial}{\partial a}+c\frac{\partial}{\partial c}\right)C_{2n+1}(a,c)b^{2n+2},\label{eq:519m}
\end{eqnarray}
\begin{equation}
M\beta^{2}/e=-\frac{1}{\pi^{2}}\sum_{n=0}^{\infty}\frac{(4n+1)!!}{(4n+2)!!}\mathscr{B}_{2n+2}C_{2n+1}(a,c)b^{2n+1},\label{eq:519n}
\end{equation}
\begin{equation}
\chi/e^{2}=-\frac{1}{\pi^{2}}\sum_{n=0}^{\infty}\frac{(4n+1)!!}{(4n)!!}\mathscr{B}_{2n+2}C_{2n+1}(a,c)b^{2n},\label{eq:517d}
\end{equation}
\begin{eqnarray}
c_{T}\beta^{3} & = & \frac{1}{2\pi^{2}}\int_{0}^{\infty}ds\int_{0}^{\infty}dt\left(12-3a\frac{\partial}{\partial a}-6c\frac{\partial}{\partial c}+ac\frac{\partial^{2}}{\partial a\partial c}+c^{2}\frac{\partial^{2}}{\partial c^{2}}\right)f(a,\sqrt{t+s^{2}+c^{2}})\nonumber \\
 &  & -\frac{1}{\pi^{2}}\sum_{n=0}^{\infty}\frac{(4n+1)!!}{(4n+4)!!}\mathscr{B}_{2n+2}\bigg[4n(4n+1)+(4n+1)a\frac{\partial}{\partial a}+(8n+2)c\frac{\partial}{\partial c}\nonumber \\
 &  & +ac\frac{\partial^{2}}{\partial a\partial c}+c^{2}\frac{\partial^{2}}{\partial c^{2}}\bigg]C_{2n+1}(a,c)b^{2n+2},\label{eq:519o}
\end{eqnarray}
where we have used the temperature factor $\beta=1/T$ and the electric
charge $e$ to make all intensive quantities dimensionless. Now all
intensive quantities are expanded at $b=0$ as a power series of $b$,
i.e. they are all analytic at $b=0$. However, for the system of massless
fermions as discussed in \citep{Fang:2020efk}, only particle number
density, entropy density and heat capacity are analytic at $b=0$,
meanwhile, other intensive quantities such as energy density, pressure,
magnetization intensity and magnetic susceptibility include a logarithmic
term $\ln b$, which are not analytic at $b=0$.

According to the asymptotic behaviors of $C_{2n+1}(a,c)$ $(n\geqslant0)$
when $c\rightarrow0$, as discussed in Appendix \ref{sec:Cn}, the
series expansions of $n$, $s$ and $c_{T}$ in Eqs. (\ref{eq:519c},
\ref{eq:519m}, \ref{eq:519o}) can return to the results of the massless
case \citep{Fang:2020efk}, which is reasonable since these intensive
quantities are analytic at $b=0$ for both massive and massless cases.

\section{Zero-field magnetic susceptibility}

\label{sec:Zero-field}
\begin{figure}[H]
\begin{centering}
\includegraphics[scale=0.5]{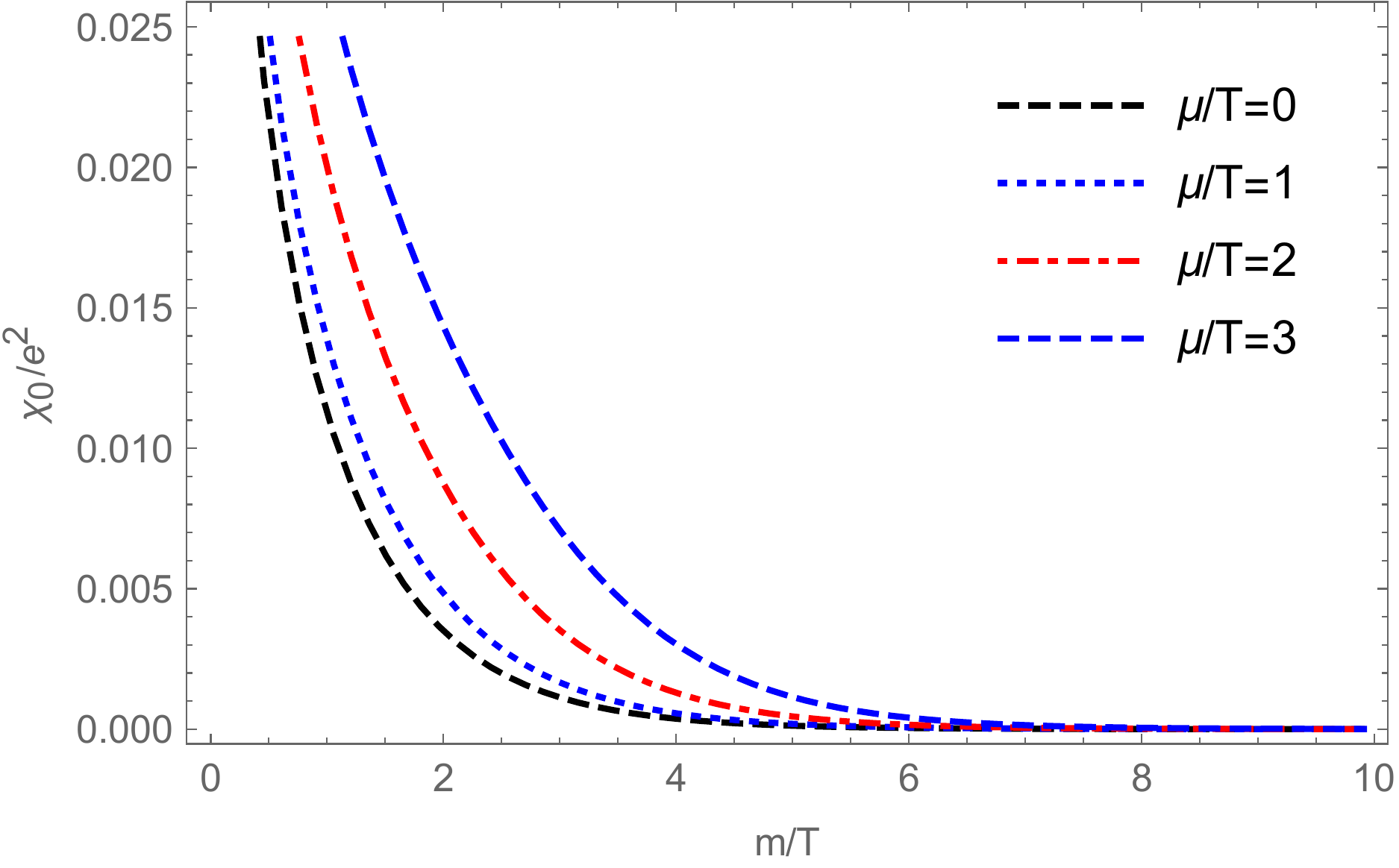}
\par\end{centering}
\caption{\label{fig:1}(color online) Mass dependence of $\chi_{0}$ at $\mu/T=0,1,2,3.$}

\end{figure}

In this section, we investigate the magnetic
susceptibility $\chi$ of the system at $B=0$, which is called zero-field
magnetic susceptibility, denoted as $\chi_{0}$. According to Eq.
(\ref{eq:517d}), zero-field magnetic susceptibility can be expressed
as
\begin{equation}
\chi_{0}(a,c)=\frac{e^{2}}{6\pi^{2}}\int_{0}^{\infty}dy\frac{1}{\sqrt{y^{2}+c^{2}}}\left(\frac{1}{e^{\sqrt{y^{2}+c^{2}}-a}+1}+\frac{1}{e^{\sqrt{y^{2}+c^{2}}+a}+1}\right).\label{eq:629a}
\end{equation}
We can see that the system of massive Dirac fermions is a paramagnetic
system since $\chi_{0}>0$. In Figure \ref{fig:1}, we plot the curves
of $\chi_{0}$ with respect to the mass parameter $c=m/T$ at $a=\mu/T=0,1,2,3$.
We find that $\chi_{0}$ is divergent at zero mass and decreases rapidly
to zero at large mass, and larger chemical potential leads to larger
$\chi_{0}$.

In Appendix \ref{sec:C1}, we obtained the expansion of $C_{1}(a,c)$
at $c=0$, so $\chi_{0}(a,c)$ can be expanded at $c=0$ as
\begin{equation}
\chi_{0}(a,c)=\frac{e^{2}}{6\pi^{2}}\left[-\frac{1}{2}\ln c^{2}+\left(D_{0}(a)+\ln2\right)+\sum_{n=1}^{\infty}\frac{(2n-1)!!}{(2n)!!}D_{n}(a)c^{2n}\right],\label{eq:517f}
\end{equation}
where $D_{n}(a)$ ($n\geqslant0$) is defined as
\begin{equation}
D_{n}(a)=-\frac{1}{(2n)!}\int_{0}^{\infty}dx\ln x\frac{d^{2n+1}}{dx^{2n+1}}\left(\frac{1}{e^{x-a}+1}+\frac{1}{e^{x+a}+1}\right).\label{eq:517e}
\end{equation}
This expansion for zero-field magnetic susceptibility is consistent
with \citep{Cangemi:1996tp}, where the imaginary time formalism and
the proper time method is used.

In the series expansion of $\chi_{0}(a,c)$ at $c=0$, besides the
regular terms $c^{2n}$, there is also a logarithmically divergent
term $\ln c^{2}$ with a negative sign. In high temperature limit,
the small mass parameter $c=m/T$ in the logarithmic term $\ln c^{2}$
will lead to a large value for $\chi_{0}$, i.e. the paramagnetic
behavior of the system becomes stronger in higher temperature. This
qualitative conclusion is consistent with recent lattice calculations
\citep{Bali:2020bcn,Buividovich:2021fsa}, where the authors considered
the system of hot and dense QCD matter.

To obtain the asymptotic behavior of $\chi_{0}$ in large mass limit,
another expression of Eq. (\ref{eq:629a}) is useful,
\begin{equation}
\chi_{0}=\frac{e^{2}}{6\pi^{2}}\int_{c}^{\infty}\frac{dx}{\sqrt{x^{2}-c^{2}}}\left(\frac{1}{e^{x-a}+1}+\frac{1}{e^{x+a}+1}\right).\label{eq:c1}
\end{equation}
Suppose that $c\gg a>0$, then Eq. (\ref{eq:c1}) becomes
\begin{equation}
\frac{\chi_{0}}{e^{2}}\sim\frac{\cosh a}{3\pi^{2}}\int_{c}^{\infty}\frac{dx}{\sqrt{x^{2}-c^{2}}}e^{-x}=\frac{\cosh a}{3\pi^{2}}K_{0}(c)\sim\frac{\cosh a}{6\pi\sqrt{c}}e^{-c},\label{eq:c2}
\end{equation}
where $K_{0}(x)$ is the virtual Hankel function and we have used
$\lim_{x\rightarrow\infty}K_{0}(x)=\frac{\pi}{2\sqrt{x}}e^{-x}$.
We find $\chi_{0}$ is exponentially suppressed at large mass. In
low temperature limit, the large mass parameter $c=m/T$ will rapidly
suppress the value of $\chi_{0}$ to be zero (always with a positive
sign). However, in recent lattice calculation \citep{Bali:2020bcn},
the diamagnetic behavior (negative susceptibility) is found at low
temperatures for QCD matter. The opposite sign of susceptibility in
low temperature may results from the absence of interaction among
particles in our calculation.

In nonrelativistic and strong degeneracy case, $\chi_{0}$ in Eq.
(\ref{eq:629a}) becomes
\begin{equation}
\chi_{0}=\frac{e^{2}}{6\pi^{2}c}\int_{0}^{\infty}dy\frac{1}{e^{\frac{y^{2}}{2c}-\frac{\xi}{T}}+1}=-\frac{e^{2}}{6\sqrt{2}\pi^{3/2}c^{1/2}}\mathrm{Li}_{1/2}\left(-e^{\xi/T}\right),\label{eq:u2}
\end{equation}
where we have defined the nonrelativistic chemical potential $\xi\equiv T(a-c)$.
Making use of $\lim_{x\rightarrow\infty}\mathrm{Li}_{1/2}\left(-e^{x}\right)=-\frac{\sqrt{x}}{\Gamma(3/2)}$,
the strong degeneracy condition with $\xi/T\gg1$ leads to
\begin{equation}
\chi_{0}=\frac{e^{2}}{3\sqrt{2}\pi^{2}}\sqrt{\frac{\xi}{m}},\label{eq:u1}
\end{equation}
which indicates that the nonrelativistic electron gas with strong
degeneracy displays weak paramagnetism \citep{Peng:2011}. In fact,
$\chi_{0}$ in Eq. (\ref{eq:u1}) can be decomposed into spin-related
and orbital angular momentum-related contributions, $\chi_{0}^{\mathrm{spin}}=\frac{3}{2}\chi_{0}$,
$\chi_{0}^{\mathrm{orbit}}=-\frac{1}{2}\chi_{0}$, which correspond
to Pauli paramagnetism and Landau diamagnetism respectively \citep{Pauli:1927,Landau:1930}.

\section{Scalar and energy-momentum tensor}

\label{sec:trace}The macroscopic thermodynamic quantities of the
system can be derived by the ensemble average of normal ordering of
corresponding field operators. In this section we will calculate the
scalar and the energy-momentum tensor of the system, which are defined
as

\begin{equation}
F=\left\langle :\overline{\psi}\psi:\right\rangle ,\label{eq:519i}
\end{equation}
\begin{equation}
T^{\mu\nu}=\frac{1}{4}\left\langle :\overline{\psi}\gamma^{\mu}iD^{\nu}\psi+\overline{\psi}\gamma^{\nu}iD^{\mu}\psi+\mathrm{h.c.}:\right\rangle ,\label{eq:519g}
\end{equation}
where $\psi$ becomes a field operator, the angular brackets means
statistical average, and the double dots enclosing the field operators
means normal ordering as adopted in \citep{Vasak:1987um,Dong:2020zci,Fang:2020efk}.

According to the rotation symmetry of the system along $z$-axis,
the nonzero components of $T^{\mu\nu}$ are $T^{00}$, $T^{11}=T^{22}$,
$T^{33}$, $T^{03}$. We have calculated energy density $\varepsilon=T^{00}$
in Eq. (\ref{eq:519d}). In the following we will calculate $T^{11}=T^{22}$,
$T^{33}$, $T^{03}$, and $F$.

The field operator $\psi(\boldsymbol{x})$ can be expanded by the
orthonormal and complete eigenfunctions in Eqs. (\ref{eq:j1}, \ref{eq:519a})
as follows,
\begin{eqnarray}
\psi(\boldsymbol{x}) & = & \sum_{k_{y},k_{z}}[a_{0}(k_{y},k_{z})\psi_{0+}(k_{y},k_{z};\boldsymbol{x})+b_{0}^{\dagger}(k_{y},k_{z})\psi_{0-}(k_{y},k_{z};\boldsymbol{x})]\nonumber \\
 &  & +\sum_{n,s,k_{y},k_{z}}[a_{ns}(k_{y},k_{z})\psi_{n+s}(k_{y},k_{z};\boldsymbol{x})+b_{ns}^{\dagger}(k_{y},k_{z})\psi_{n-s}(k_{y},k_{z};\boldsymbol{x})],\label{eq:519h}
\end{eqnarray}
where $a_{n}$, $b_{n}$, $a_{n}^{\dagger}$, $b_{n}^{\dagger}$ are
annihilation and creation operators for fermions and antifermions.
As calculated in \citep{Dong:2020zci} , the ensemble average of normal
ordering of $a_{n}^{\dagger}a_{n}$ and $b_{n}b_{n}^{\dagger}$ are
\begin{eqnarray}
\left\langle :a_{ns}^{\dagger}(k_{y},k_{z})a_{ns}(k_{y},k_{z}):\right\rangle  & = & \frac{1}{e^{\beta(E_{n}-\mu)}+1},\label{eq:57b}\\
\left\langle :b_{ns}(k_{y},k_{z})b_{ns}^{\dagger}(k_{y},k_{z}):\right\rangle  & = & -\frac{1}{e^{\beta(E_{n}+\mu)}+1}.\label{eq:54k}
\end{eqnarray}
Plugging Eq. (\ref{eq:519h}) into Eqs. (\ref{eq:519i}-\ref{eq:519g})
and making use of Eqs. (\ref{eq:57b}), (\ref{eq:54k}), one can obtain
$T^{03}=0$ and
\begin{equation}
T^{11}=T^{22}=\frac{(eB)^{2}}{\pi^{2}}\sum_{n=1}^{\infty}\int_{0}^{\infty}dk_{z}\frac{n}{E_{n}}\left(\frac{1}{e^{\beta(E_{n}-\mu)}+1}+\frac{1}{e^{\beta(E_{n}+\mu)}+1}\right),\label{eq:71u}
\end{equation}
\begin{equation}
T^{33}=\frac{eB}{\pi^{2}}\sum_{n=0}^{\infty}\frac{1}{1+\delta_{n,0}}\int_{0}^{\infty}dk_{z}\frac{k_{z}^{2}}{E_{n}}\left(\frac{1}{e^{\beta(E_{n}-\mu)}+1}+\frac{1}{e^{\beta(E_{n}+\mu)}+1}\right),\label{eq:71v}
\end{equation}
\begin{equation}
F=\frac{eB}{\pi^{2}}\sum_{n=0}^{\infty}\frac{1}{1+\delta_{n,0}}\int_{0}^{\infty}dk_{z}\frac{m}{E_{n}}\left(\frac{1}{e^{\beta(E_{n}-\mu)}+1}+\frac{1}{e^{\beta(E_{n}+\mu)}+1}\right).\label{eq:a2-1}
\end{equation}
Due to the absence of axial chemical potential $\mu_{5}$ in our formulism,
$T^{03}$ vanishes \citep{Gao:2012ix,Yang:2020mtz}. Further calculations
of $T^{11}$, $T^{22},$ $T^{33}$ and $F$ give
\begin{equation}
T^{11}=T^{22}=\frac{1}{\beta^{4}}\left(1-b\frac{\partial}{\partial b}\right)g(a,b,c),\label{eq:71w}
\end{equation}
\begin{equation}
T^{33}=\frac{1}{\beta^{4}}g(a,b,c)=p,\label{eq:71x}
\end{equation}
\begin{equation}
F=-\frac{1}{\beta^{3}}\frac{\partial}{\partial c}g(a,b,c).\label{eq:71y}
\end{equation}
We find that $T^{33}$ is just the pressure $p$.

According to Dirac equation $\left(i\gamma^{\mu}D_{\mu}-m\right)\psi=0$,
the trace of $T^{\mu\nu}$ is
\begin{equation}
g_{\mu\nu}T^{\mu\nu}=T^{00}-T^{11}-T^{22}-T^{33}=mF.\label{eq:t1}
\end{equation}
The expressions of $T^{00}$, $T^{11}$, $T^{22}$, $T^{33}$, $F$
in Eqs. (\ref{eq:512j}, \ref{eq:71w}-\ref{eq:71y}) automatically
satisfy Eq. (\ref{eq:t1}). Now we calculate the massless limit of
Eq. (\ref{eq:t1}). From the series expansion of $g(a,b,c)$ and $F$
in Eqs. (\ref{eq:519k}, \ref{eq:71y}), the trace equation becomes
\begin{eqnarray}
g_{\mu\nu}T^{\mu\nu} & = & \frac{c}{\beta^{4}}\bigg[-\frac{1}{2\pi^{2}}\int_{0}^{\infty}ds\int_{0}^{\infty}dt\frac{\partial}{\partial c}f\left(a,\sqrt{t+s^{2}+c^{2}}\right)\nonumber \\
 &  & +\frac{1}{\pi^{2}}\sum_{n=0}^{\infty}\frac{(4n+1)!!}{(4n+4)!!}\mathscr{B}_{2n+2}\frac{\partial}{\partial c}C_{2n+1}(a,c)b^{2n+2}\bigg].\label{eq:d1}
\end{eqnarray}
The first integration in Eq. (\ref{eq:d1}) tends to be zero as $c\rightarrow0$.
Making use of the asymptotic behaviors of $C_{n}(a,c)$ at $c=0$
in Appendix \ref{sec:Cn}, one can obtain
\begin{equation}
\lim_{c\rightarrow0}c\frac{\partial}{\partial c}C_{n}(a,c)=\delta_{n,1}.\label{eq:s2}
\end{equation}
Then Eq. (\ref{eq:d1}) gives
\begin{equation}
\lim_{c\rightarrow0}g_{\mu\nu}T^{\mu\nu}=\frac{e^{2}B^{2}}{12\pi^{2}},\label{eq:s3}
\end{equation}
which is just the trace anomaly equation \citep{Peskin:1995,Yang:2020mtz}
in the case of a pure magnetic field with the field strength invariant
$F^{\mu\nu}F_{\mu\nu}=2B^{2}$.

\section{Mass correction to chiral separation effect}

\label{sec:CSE}Similar to the calculation for $T^{\mu\nu}$ in Sec.
\ref{sec:trace}, the vector and axial currents $J_{V}^{\mu}$, $J_{A}^{\mu}$
can also be derived through ensemble average, i.e.
\begin{equation}
J_{V}^{\mu}=\left\langle :\overline{\psi}\gamma^{\mu}\psi:\right\rangle ,\label{eq:519f-1}
\end{equation}

\begin{equation}
J_{A}^{\mu}=\left\langle :\overline{\psi}\gamma^{\mu}\gamma^{5}\psi:\right\rangle .\label{eq:519f}
\end{equation}
The zero component of $J_{V}^{\mu}$ is just the particle number density
calculated in Eq. (\ref{eq:519c}). Plugging Eq. (\ref{eq:519h})
into Eqs. (\ref{eq:519f-1}, \ref{eq:519f}) and making use of Eqs. (\ref{eq:57b}, \ref{eq:54k})
one can obtain

\begin{equation}
J_{V}^{i}=J_{A}^{k}=J_{A}^{0}=0,\ \ \ (i=1,2,3;\ k=1,2),\label{eq:71s}
\end{equation}
\begin{equation}
J_{A}^{3}=\frac{eB}{2\pi^{2}}\int_{0}^{\infty}dk_{z}\left(\frac{1}{e^{\beta(\sqrt{m^{2}+k_{z}^{2}}-\mu)}+1}-\frac{1}{e^{\beta(\sqrt{m^{2}+k_{z}^{2}}+\mu)}+1}\right).\label{eq:71t}
\end{equation}
Due to the absence of axial chemical potential $\mu_{5}$ in our formulism,
$J_{V}^{i}$ and $J_{A}^{0}$ vanish \citep{Gao:2012ix,Sheng:2017lfu,Yang:2020mtz}.
Fluctuation and dissipation of axial charge $J_{A}^{0}$ from massive
quarks is discussed by one of the authors \citep{Hou:2017szz}. The
nonzero component $J_{A}^{3}$ is linear for the magnetic field $B$,
which is the chiral separation effect (CSE) for the system of massive
Dirac fermions \citep{Gorbar:2013upa,Lin:2018aon}.

If we take $m=0$, the coefficient of $eB$ in Eq. (\ref{eq:71t})
can be integrated out,
\begin{equation}
J_{A}^{3}=\frac{eB\mu}{2\pi^{2}}.\label{eq:73a}
\end{equation}
We know that, for massless fermion system, the dynamics of right-
and left-handed fermions decouples, so we can introduce the particle
number currents $J_{R}^{\mu}$, $J_{L}^{\mu}$ for right- and left-handed
fermion systems, respectively. Now the vector and axial currents $J_{V}^{\mu}$,
$J_{A}^{\mu}$ can be expressed as
\begin{equation}
J_{V}^{\mu}=J_{R}^{\mu}+J_{L}^{\mu},\ \ \ \ J_{A}^{\mu}=J_{R}^{\mu}-J_{L}^{\mu}.\label{eq:73b}
\end{equation}
Since we have set $\mu_{5}=0$, then chemical potentials $\mu_{R}$,
$\mu_{L}$ of right- and left-handed fermions are equal. Combining
Eq. (\ref{eq:71s}) and Eq. (\ref{eq:73b}), one can obtain
\begin{equation}
J_{R}^{3}=\frac{eB\mu_{R}}{4\pi^{2}},\ \ \ \ J_{L}^{3}=-\frac{eB\mu_{L}}{4\pi^{2}},\label{eq:73c}
\end{equation}
which are the chiral magnetic effects for right- and left-handed fermion
systems respectively \citep{Fukushima:2008xe,Kharzeev:2010gr}.

We rewrite Eq. (\ref{eq:71t}) as $J_{A}^{3}=B\sigma_{B}$, and $\sigma_{B}$
is the CSE coefficient,
\begin{equation}
\sigma_{B}=\frac{eT}{2\pi^{2}}\int_{0}^{\infty}dx\left(\frac{1}{e^{\sqrt{x^{2}+c^{2}}-a}+1}-\frac{1}{e^{\sqrt{x^{2}+c^{2}}+a}+1}\right),\label{eq:w1}
\end{equation}
which is consistent with \citep{Lin:2018aon}, where the authors showed
that the presence of mass generically suppresses the CSE coefficient
with less suppression at larger chemical potential.

At zero temperature, Eq. (\ref{eq:w1}) gives
\begin{equation}
\sigma_{B}=\frac{e}{2\pi^{2}}\sqrt{\mu^{2}-m^{2}}\left[\theta(\mu-m)-\theta(-\mu-m)\right],\label{eq:x1}
\end{equation}
which vanishes for $\mu^{2}<m^{2}$, and returns to $\frac{e\mu}{2\pi^{2}}$
in massless limit. This agrees with the free case \citep{Gorbar:2013upa,Lin:2018aon}.

At nonzero temperature, we will expand the the CSE coefficient $\sigma_{B}$
at $m=0$. Another expression of Eq. (\ref{eq:w1}) is useful,
\begin{equation}
\sigma_{B}=\frac{eT}{2\pi^{2}}\frac{d}{da}F(a,c),\label{eq:w2}
\end{equation}
\begin{equation}
F(a,c)=\int_{|c|}^{\infty}dy\frac{y}{\sqrt{y^{2}-c^{2}}}\left[\ln\left(1+e^{a-y}\right)+\ln\left(1+e^{-a-y}\right)\right].\label{eq:w3}
\end{equation}
In the recent article by some of us \citep{Fang:2020efk}, Eqs. (B3,
B12, B14) can give
\begin{equation}
\sigma_{B}=\frac{eT}{2\pi^{2}}\sum_{n=0}^{\infty}\frac{(2n-1)!!}{(2n)!!}E_{n}(a)\left(\frac{m}{T}\right)^{2n},\label{eq:w4}
\end{equation}
where $E_{n}(a)$ are defined as
\begin{equation}
E_{0}(a)=a,\label{eq:w5}
\end{equation}
\begin{equation}
E_{n}(a)=\frac{1}{(2n-1)!}\frac{d^{2n}}{da^{2n}}\left\{ \frac{d}{ds}\left[\mathrm{Li}_{s}(-e^{a})-\mathrm{Li}_{s}(-e^{-a})\right]\right\} _{s=1},\ \ (n\geqslant1),\label{eq:w6}
\end{equation}
and $\mathrm{Li}_{s}(z)$ is the polylogarithm function. The $n=1$
term in Eq. (\ref{eq:w4}) is consistent with \citep{Lin:2018aon}.
There is no logarithmic term $\ln(m/T)$ in the expansion of $\sigma_{B}$,
which is different from the case of zero-field magnetic susceptibility
discussed in Sec. \ref{sec:Zero-field}. In the recent article \citep{Fang:2020efk},
Eqs. (D9, C9) can give following asymptotic behaviors of $E_{n}(a)$
at small and lager chemical potential respectively,
\begin{equation}
\lim_{a\rightarrow0}E_{n}(a)=(-1)^{n}\frac{2n}{\pi^{2n}}\zeta(2n+1)\left(2-2^{-2n}\right)a,\label{eq:w7}
\end{equation}
\begin{equation}
\lim_{a\rightarrow\infty}E_{n}(a)=-\frac{1}{2n-1}\frac{1}{a^{2n-1}}.\label{eq:w8}
\end{equation}
Making use of $\lim_{s\rightarrow0}s\zeta(s+1)=1$, we find that Eqs.
(\ref{eq:w7}, \ref{eq:w8}) both apply to $n=0$ case. The asymptotic
behavior in Eq. (\ref{eq:w8}) implies that mass corrections to $\sigma_{B}$
at high orders become smaller at larger chemical potential.

\section{Summary}

\label{sec:Summary}In this article we have studied the influence
of the magnetic field on the system of massive Dirac fermions. From
the Landau levels for a single massive Dirac fermion in a uniform
magnetic field, we construct the partition function for the system,
through which all thermodynamic intensive quantities can be derived.
Making use of Abel-Plana formula, all thermodynamic quantities can
be expanded as power series of the dimensionless variable $b=2eB/T^{2}$,
i.e. they are all analytic at $b=0$, which is different from the
massless case. We expand the zero-field magnetic susceptibility $\chi_{0}$
at $m=0$, and a logarithmic term appears at leading order. The asymptotic
behavior of $\chi_{0}$ at high temperature is qualitatively consistent
with recent lattice calculations. In nonrelativistic limit, $\chi_{0}$
can return to the result of electron gas with strong degeneracy. The
macroscopic thermodynamic quantities, such as scalar, vector current,
axial vector current and energy-momentum tensor, can be obtained by
the ensemble average of normal ordering of corresponding field operators.
For massless chiral fermions, our results recover the chiral magnetic
effect for right- and left-handed fermions. We discuss mass correction
to chiral separation effect. The presence of mass exponentially suppresses
the CSE coefficient, and we expand the CSE coefficient at $m=0$ to
any order.

\section{\textup{Acknowledgments}}

We thank Hai-Cang Ren and Xin-Li Sheng for helpful discussions. This work
is supported by the National Natural Science Foundation of China
(Grant Nos. 11890713, 11735007, 11890711, and 11947228), and the 
Chinese Postdoctoral Science Foundation (Grant No. 2019M662316).

\appendix

\section{Expansion of $g(a,b,c)$ at $b=0$}

\label{sec:Expansion-app} Making use of Abel-Plana formula, the grand
partition function can be expressed as
\begin{equation}
g(a,b,c)=\frac{1}{2\pi^{2}}\int_{0}^{\infty}ds\int_{0}^{\infty}dtf(a,\sqrt{t+s^{2}+c^{2}})+\frac{b}{2\pi^{2}}\times i\int_{0}^{\infty}dt\frac{F(a,c,\sqrt{itb})-F(a,c,\sqrt{-itb})}{e^{2\pi t}-1}.\label{eq:517c}
\end{equation}
where $a=\mu\beta$, $b=2eB\beta^{2}$, $c=m\beta$ and $f(a,x)$,
$F(a,c,x)$ are defined as
\begin{equation}
f(a,x)=\ln(1+e^{a-x})+\ln(1+e^{-a-x}),\label{eq:72b}
\end{equation}

\begin{equation}
F(a,c,x)=\int_{0}^{\infty}dsf(a,\sqrt{x^{2}+s^{2}+c^{2}}).\label{eq:517b}
\end{equation}
By the variable transformation $y=\sqrt{x^{2}+s^{2}}$ in Eq. (\ref{eq:517b}),
$F(a,c,x)$ can be rewritten as
\begin{equation}
F(a,c,x)=\int_{|x|}^{\infty}dy\frac{y}{\sqrt{y^{2}-x^{2}}}f(a,\sqrt{y^{2}+c^{2}}).\label{eq:72f}
\end{equation}
Making use of following Taylor expansion,
\begin{equation}
\frac{y}{\sqrt{y^{2}-x^{2}}}=\sum_{n=0}^{\infty}\frac{(2n-1)!!}{(2n)!!}\frac{x^{2n}}{y^{2n}},\label{eq:ap4}
\end{equation}
 $F(a,c,x)$ becomes
\begin{equation}
F(a,c,x)=\sum_{n=0}^{\infty}\frac{(2n-1)!!}{(2n)!!}x^{2n}d_{n}(a,c,x),\label{eq:z1}
\end{equation}
where we have defined $d_{n}(a,c,x)$ as

\begin{equation}
d_{n}(a,c,x)=\int_{|x|}^{\infty}dy\frac{1}{y^{2n}}f(a,\sqrt{y^{2}+c^{2}}).\label{eq:510w}
\end{equation}
Since $d_{n}(a,c,x)=d_{n}(a,c,-x)$, the derivative of $d_{n}(a,c,x)$
with respect to $x$ is

\begin{equation}
d_{n}^{\prime}(a,c,x)=-\frac{|x|}{x^{2n+1}}f(a,\sqrt{x^{2}+c^{2}}).\label{eq:72h}
\end{equation}
We can expand $f(a,\sqrt{x^{2}+c^{2}})$ at $x=0$ as follows,
\begin{equation}
f(a,\sqrt{x^{2}+c^{2}})=\sum_{k=0}^{\infty}w_{2k}(a,c)x^{2k},\label{eq:72i}
\end{equation}
then $d_{n}(a,c,x)$ becomes
\begin{equation}
d_{n}(a,c,x)=|x|\sum_{k=0}^{\infty}w_{2k}(a,c)\frac{1}{2n-2k-1}x^{2k-2n}+C_{n}(a,c),\label{eq:72j}
\end{equation}
where $C_{n}(a,c)$ is independent of $x$. Now $F(a,c,x)$ in Eq.
(\ref{eq:z1}) becomes
\begin{equation}
F(a,c,x)=\sum_{n=0}^{\infty}\frac{(2n-1)!!}{(2n)!!}C_{n}(a,c)x^{2n},\label{eq:517b-1}
\end{equation}
where we have used
\begin{equation}
\sum_{n=0}^{\infty}\frac{(2n-1)!!}{(2n)!!}\frac{1}{2n-2k-1}=0,\ \ (k=0,1,2,\cdots).\label{eq:72k}
\end{equation}

The coefficient $C_{n}(a,c)$ in Eq. (\ref{eq:517b-1}) is the constant
term when $d_{n}(a,c,x)$ in expanded at $x=0$. When $n>0$, we can
rewrite $d_{n}(a,c,x)$ in (\ref{eq:510w}) through integration by
parts as
\begin{eqnarray}
d_{n}(a,c,x) & = & \sum_{k=0}^{2n-2}\frac{(2n-k-2)!}{(2n-1)!}\frac{1}{x^{2n-k-1}}\frac{d^{k}}{dx^{k}}f(a,\sqrt{x^{2}+c^{2}})-\frac{\ln x}{(2n-1)!}\frac{d^{2n-1}}{dx^{2n-1}}f(a,\sqrt{x^{2}+c^{2}})\nonumber \\
 &  & -\frac{1}{(2n-1)!}\int_{x}^{\infty}dy\ln y\frac{d^{2n}}{dy^{2n}}f(a,\sqrt{y^{2}+c^{2}}).\label{eq:510a2}
\end{eqnarray}
which implies
\begin{equation}
C_{n}(a,c)=-\frac{1}{(2n-1)!}\int_{0}^{\infty}dy\ln y\frac{d^{2n}}{dy^{2n}}f(a,\sqrt{y^{2}+c^{2}}).\label{eq:517h}
\end{equation}
When $n=0$, we have
\begin{equation}
C_{0}(a,c)=\int_{0}^{\infty}dyf(a,\sqrt{y^{2}+c^{2}}).\label{eq:72l}
\end{equation}

Substituting Eq. (\ref{eq:517b-1}) into Eq. (\ref{eq:517c}) gives

\begin{equation}
g(a,b,c)=\frac{1}{2\pi^{2}}\int_{0}^{\infty}ds\int_{0}^{\infty}dtf(a,\sqrt{t+s^{2}+c^{2}})-\frac{1}{\pi^{2}}\sum_{n=0}^{\infty}\frac{(4n+1)!!}{(4n+4)!!}\mathscr{B}_{2n+2}C_{2n+1}(a,c)b^{2n+2},\label{eq:72m}
\end{equation}
where we have used following integrations,
\begin{equation}
\int_{0}^{\infty}dt\frac{t^{2n+1}}{e^{2\pi t}-1}=(-1)^{n}\frac{\mathscr{B}_{2n+2}}{4n+4},\ \ (n\geqslant0),\label{eq:510a4}
\end{equation}
with Bernoulli numbers $\mathscr{B}_{n}$ defined as
\begin{equation}
\frac{t}{e^{t}-1}=\sum_{n=0}^{\infty}\frac{\mathscr{B}_{n}}{n!}t^{n}.\label{eq:510a5}
\end{equation}

\section{Asymptotic behaviors of $C_{n}(a,c)$ $(n\geqslant1)$ at $c=0$}

\label{sec:Cn}The expression of $C_{n}(a,c)$ $(n\geqslant1)$ in
Eq. (\ref{eq:517h}) is

\begin{equation}
C_{n}(a,c)=-\frac{1}{(2n-1)!}\int_{0}^{\infty}dy\ln y\frac{d^{2n}}{dy^{2n}}f(a,\sqrt{y^{2}+c^{2}}),\label{eq:0907}
\end{equation}
where $f(a,x)=\ln(1+e^{a-x})+\ln(1+e^{-a-x})$. Making use of integration
by parts, Eq. (\ref{eq:0907}) becomes
\begin{equation}
C_{n}(a,c)=-\frac{1}{(2n-1)!}\int_{0}^{\infty}dy\frac{1}{y}\frac{d^{2n-2}}{dy^{2n-2}}\left[\left(\frac{1}{e^{\sqrt{y^{2}+c^{2}}-a}+1}+\frac{1}{e^{\sqrt{y^{2}+c^{2}}+a}+1}\right)\frac{y}{\sqrt{y^{2}+c^{2}}}\right],\label{eq:0907b}
\end{equation}
where the second equal sign is valid for $n>1$ with an arbitrary
$c$, and for $n=1$ with $c\neq0$.

When $c=0$, Eq. (\ref{eq:0907b}) becomes
\begin{equation}
C_{n}(a,0)=-\frac{1}{(2n-1)!}\int_{0}^{\infty}dy\frac{1}{y}\frac{d^{2n-2}}{dy^{2n-2}}\left(\frac{1}{e^{y-a}+1}+\frac{1}{e^{y+a}+1}\right).\label{eq:0907d}
\end{equation}
The expression in the bracket in the integrand can be expanded at
$y=0$ as follows,
\begin{equation}
\frac{1}{e^{y-a}+1}+\frac{1}{e^{y+a}+1}=1+\#y+\#y^{3}+\#y^{5}+\cdots,\label{eq:0907c}
\end{equation}
with ``$\#$'' representing some coefficients independent of $y$.
The expansion in Eq. (\ref{eq:0907c}) implies that, $C_{n}(a,0)$
is divergent for $n=1,$ and convergent for $n>1$. In Sec. \ref{sec:C1},
we have proved that, the asymptotic behavior of $C_{1}(a,c)$ at $c=0$
is
\begin{equation}
C_{1}(a,c)=\frac{1}{2}\ln c^{2}+(c^{2k}\ \mathrm{terms\ with\ }k\geqslant0).\label{eq:b1}
\end{equation}

Now we analyse the asymptotic behaviors of $C_{n}(a,c)$ at $c=0$
for $n>1$. The derivative of $C_{n}(a,c)$ with respect to $c^{2}$
is
\begin{eqnarray}
\frac{d}{dc^{2}}C_{n}(a,c) & = & \frac{1}{(2n-1)!}\int_{0}^{\infty}dy\ln y\frac{d^{2n}}{dy^{2n}}\left[\frac{1}{2\sqrt{y^{2}+c^{2}}}\left(\frac{1}{e^{\sqrt{y^{2}+c^{2}}-a}+1}+\frac{1}{e^{\sqrt{y^{2}+c^{2}}+a}+1}\right)\right].\nonumber \\
\label{eq:a1}
\end{eqnarray}
When $c=0$, the asymptotic behavior of the expression in the square
brackets in Eq. (\ref{eq:a1}) at $y=0$ is $1/y$, then the integration
in Eq. (\ref{eq:a1}) is divergent. Since $C_{n}(a,0)$ is finite,
we can obtain
\begin{equation}
C_{n}(a,c)=\#c^{2}\ln c^{2}+(c^{2k}\ \mathrm{terms\ with\ }k\geqslant0).\label{eq:b3}
\end{equation}

\section{Expansion of $C_{1}(a,c)$ at $c=0$}

\label{sec:C1}When $n=1$, Eq. (\ref{eq:0907b}) becomes
\begin{eqnarray}
C_{1}(a,c) & = & -\int_{0}^{\infty}dy\frac{1}{\sqrt{y^{2}+c^{2}}}\left(\frac{1}{e^{\sqrt{y^{2}+c^{2}}-a}+1}+\frac{1}{e^{\sqrt{y^{2}+c^{2}}+a}+1}\right)\nonumber \\
 & = & -\int_{|c|}^{\infty}dx\frac{1}{x}\left(1-\frac{c^{2}}{x^{2}}\right)^{-\frac{1}{2}}\left(\frac{1}{e^{x-a}+1}+\frac{1}{e^{x+a}+1}\right).\label{eq:517i}
\end{eqnarray}
Since $x^{2}>c^{2}$, we can use the Taylor expansion for $(1-c^{2}/x^{2})^{-1/2}$
in Eq. (\ref{eq:ap4}), which leads to
\begin{equation}
C_{1}(a,c)=-\sum_{n=0}^{\infty}\frac{(2n-1)!!}{(2n)!!}c^{2n}X_{n}(a,c),\label{eq:72n}
\end{equation}
where $X_{n}(a,c)$ can be written as follows,

\begin{eqnarray}
X_{n}(a,c) & = & \int_{|c|}^{\infty}dx\left[\frac{1}{x^{2n+1}}\left(\frac{1}{e^{x-a}+1}+\frac{1}{e^{x+a}+1}-1\right)+\frac{1}{x^{2n+1}}\right].\label{eq:72o}
\end{eqnarray}

The derivative of $X_{n}(a,c)$ with respect to $c$ is
\begin{equation}
X_{n}^{\prime}(a,c)=-\frac{|c|}{c^{2n+2}}\left(\frac{1}{e^{c-a}+1}+\frac{1}{e^{c+a}+1}-1\right)-\frac{1}{c^{2n+1}}.\label{eq:517j}
\end{equation}
The term in the bracket in Eq. (\ref{eq:517j}) is an odd function
of $c$ which can be expanded at $c=0$ as
\begin{equation}
\frac{1}{e^{c-a}+1}+\frac{1}{e^{c+a}+1}-1=\sum_{k=0}^{\infty}v_{2k+1}(a)c^{2k+1},\label{eq:72p}
\end{equation}
then $X_{n}(a,c)$ can be obtained from $X_{n}^{\prime}(a,c)$,

\begin{equation}
X_{n}(a,c)=-|c|\sum_{k=0}^{\infty}v_{2k+1}(a)\frac{1}{2k-2n+1}c^{2k-2n}+\left\{ \begin{array}{cc}
D_{0}(a)-\frac{1}{2}\ln c^{2}, & n=0\\
D_{n}(a)+\frac{1}{2n}c^{-2n}, & n>0
\end{array}\right.,\label{eq:72q}
\end{equation}
where $D_{n}(a)$ are independent of $c$ and can be determined by
the same method as the calculation of $C_{n}(a,c)$ in Sec. \ref{sec:Expansion-app}.
The result of $D_{n}(a)$ is

\begin{equation}
D_{n}(a)=-\frac{1}{(2n)!}\int_{0}^{\infty}dx\ln x\frac{d^{2n+1}}{dx^{2n+1}}\left(\frac{1}{e^{x-a}+1}+\frac{1}{e^{x+a}+1}\right).\label{eq:72r}
\end{equation}
Making use of
\begin{equation}
\sum_{n=0}^{\infty}\frac{(2n-1)!!}{(2n)!!}\frac{1}{2n-2k-1}=0,\ \ (k=0,1,2,\cdots),\label{eq:72s}
\end{equation}
\begin{equation}
\sum_{n=1}^{\infty}\frac{(2n-1)!!}{(2n)!!}\frac{1}{n}=\ln4,\label{eq:72t}
\end{equation}
we get
\begin{equation}
C_{1}(a,c)=\frac{1}{2}\ln c^{2}-\left[D_{0}(a)+\ln2\right]-\sum_{n=1}^{\infty}\frac{(2n-1)!!}{(2n)!!}D_{n}(a)c^{2n}.\label{eq:517k}
\end{equation}

According to the Appendix D in \citep{Fang:2020efk}, $D_{n}(a)$
can be expanded at $a=0$ as follows,
\begin{equation}
D_{n}(a)=(-\ln4-\gamma)\delta_{n,0}-\frac{2}{(2n)!}\sum_{k=0}^{\infty}\left(2^{2n+2k+1}-1\right)\zeta^{\prime}(-2n-2k)\frac{a^{2k}}{(2k)!}.\label{eq:72u}
\end{equation}

\bibliographystyle{apsrev}
\addcontentsline{toc}{section}{\refname}\bibliography{ref-20200610}

\end{document}